\documentclass{article}
\usepackage{spconf,amsmath,graphicx,hyperref}
\usepackage{xcolor}
\usepackage{amssymb}
\usepackage{multirow}
\usepackage{booktabs}
\usepackage{array}
\usepackage{algorithmic}
\usepackage{acronym}
\usepackage{float}  
\usepackage{indentfirst}
\usepackage{graphicx}
\usepackage{subcaption}
\usepackage{xcolor}
\usepackage{cite}

\newcommand{\mymodel}{S-PRESSO}
\acrodef{RVQ}{Residual Vector Quantization} 
\acrodef{DiT}{Diffusion Transformer}

\title{\mymodel: Ultra low bitrate sound effect compression with diffusion autoencoders and offline quantization}

\name{Zineb Lahrichi${}^\ddag$${}^*$, Gaëtan Hadjeres${}^\ddag$, Gaël Richard${}^*$, Geoffroy Peeters${}^*$}
\address{${}^\ddag$Sony AI, ${}^*$LTCI, Télécom Paris, Institut Polytechnique de Paris}

\begin{document}
\ninept
\maketitle

\begin{abstract}

Neural audio compression models have recently achieved extreme compression rates, enabling efficient latent generative modeling. Conversely, latent generative models have been applied to compression, pushing the limits of continuous and discrete approaches. However, existing methods remain constrained to low-resolution audio and degrade substantially at very low bitrates, where audible artifacts are prominent. In this paper, we present \textbf{\mymodel}, a 48kHz sound effect compression model that produces both continuous and discrete embeddings at \textit{ultra-low bitrates}, down to 0.096~kbps, via offline quantization.
Our model relies on a pretrained latent diffusion model to decode compressed audio embeddings learned by a latent encoder. Leveraging the generative priors of the diffusion decoder, we achieve \textit{extremely low frame rates}, down to 1Hz (750x compression rate), producing convincing and realistic reconstructions at the cost of exact fidelity. 
Despite operating at high compression rates, we demonstrate that \mymodel{} outperforms both continuous and discrete baselines in audio quality, acoustic similarity and reconstruction metrics. Audio samples are available at \url{https://zineblahrichi.github.io/s-presso/}

\end{abstract}


\section{Introduction}
\label{sec:intro}

In recent years, substantial efforts have been devoted to designing low to ultra-low bitrate codecs, motivated by practical deployment of efficient codecs and the creation of compact representations suitable for latent generative models.
However, as reported in the image domain \cite{careil2024imagecompressionperfectrealism}, neural compression methods are typically optimized for the rate-distortion trade-off at the expense of perceptual quality and realism, often producing artifacts and less natural images at low bitrates. 
A similar limitation holds for audio: codecs built on residual vector quantization and adversarial training (RVQ-GANs) \cite{zeghidour2021soundstream, défossez2022highfidelityneuralaudio, kumar2023highfidelityaudiocompressionimproved} are deterministic and target exact reconstruction, but at high compression introduce audible degradations such as metallic or robotic timbres. Despite adversarial objectives, their perceptual quality remains fundamentally constrained by the bitrate.

To address these limitations, generative models offer a promising alternative, leveraging their strong generative priors to shift the focus from a bitrate/quality trade-off to a bitrate/acoustic similarity trade-off. Here, acoustic similarirty refers to the perceptual resemblance of two sounds as originating from the same source with comparable characteristics over time. While strict similarity can be critical for certain applications (e.g.,
lossless music streaming), this is less true in dynamic environments such as video games.

Moreover, the stochasticity of generative models can even be advantageous, providing natural variations that prevents repetitive playback of audio samples. For example, avoiding identical footstep sounds when a character walks in a video game helps to enhance perceptual realism.

Ultra–low bitrate codecs using generative models were initially developed for speech \cite{siahkoohi2022ultralowbitratespeechcodingpretrained} and have since been extended to general audio and music \cite{Liu_2024, xu2025mucodecultralowbitratemusic}, achieving bitrates of only a few hundred bits per second. However, to our knowledge, these approaches remain limited in bandwidth ($<$ 32 kHz) and exhibit noticeable quality degradation at very low bitrates.

In this paper, we make further progress towards ultra low bitrate compression of high quality audio, focusing on sound effects. We introduce \textbf{\mymodel{}}, a diffusion autoencoder that relies on a pretrained latent diffusion model to decode compressed audio embeddings learned by a latent compressor.

In order to encode both \textit{continuous} and \textit{discrete} embeddings, we adopt a three-step training procedure as in \cite{lahrichi2025qincodecneuralaudiocompression}. 
This includes (i) learning compressed representations via continuous diffusion autoencoder training, (ii) offline neural quantization, and (iii) diffusion decoder finetuning, which enables compact yet expressive representations. The proposed model achieves compression ratios up to 750x on 48kHz audio, producing discrete representations at bitrates as low as 0.096kbps while preserving perceptual quality. Finally, leveraging diffusion model priors, our method outperforms strong continuous and discrete baselines in sound quality and acoustic similarity, as corroborated by human evaluations, delivering realistic and high-quality reconstructions even at ultra-low frame rates (down to 1Hz).

\begin{figure*}[t!] 
    \centering
    \includegraphics[width=0.90\textwidth]{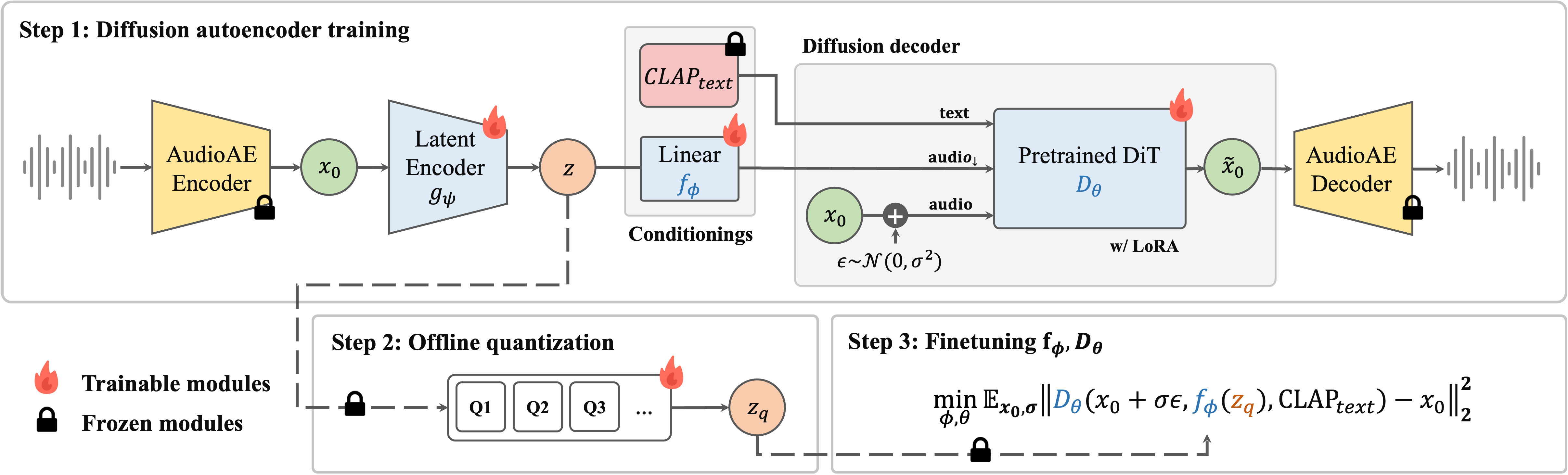}
    \caption{Overview of our method. \textbf{Step 1}: An audio clip is encoded into latent vectors $x_0$ by a
    low-compression audio autoencoder. It is then compressed into latents $z$, which are upsampled by $f_\phi$ to condition the decoder $D_\theta$, a \ac{DiT} pretrained to reconstruct $x_0$ from noised inputs. $D_\theta$ is finetuned using LoRA adapters, jointly trained with the latent encoder $g_{\psi}$ and $f_\phi$. \textbf{Step 2}: The features $z$ are then quantized offline into $z_q$. \textbf{Step 3}: the diffusion decoder $D_\theta$ is finetuned on $z_q$, to compensate for quantization-induced degradation.}
    \label{fig:overview}
\end{figure*}

\section{Related Works: Low Bitrate Codecs}
\label{sec:related works}

Neural audio compression has recently advanced beyond traditional codecs in rate/distortion trade-offs. RVQ-GAN models~\cite{zeghidour2021soundstream, défossez2022highfidelityneuralaudio, kumar2023highfidelityaudiocompressionimproved} achieve high-fidelity reconstructions at moderate bitrates, but reconstruction quality typically collapses below $\sim$3kbps. Alternative strategies lower the bitrate by reducing frame rates rather than improving quantization~\cite{casanova2025low, li2025dualcodec}, yet these remain tailored to speech and narrowband signals.

Advances in generative modeling have enabled compression at much lower rates. Early works applied WaveNet decoders \cite{kleijn2018wavenet} to speech, achieving rates of $\sim$ 2kbps \cite{siahkoohi2022ultralowbitratespeechcodingpretrained}, but the limited receptive fields of WaveNets restricted long-range temporal modeling. To address this, these approaches were further improved by using transformers with GAN decoders, capturing longer dependencies and pushing speech compression down to 600 bps \cite{siahkoohi2022ultralowbitratespeechcodingpretrained}. 

Extensions to general audio and music leverage pretrained semantic or acoustic latent spaces decoded by diffusion models~\cite{Liu_2024, xu2025mucodecultralowbitratemusic}, achieving extreme compression down to a few hundred bps. However, these methods remain constrained to narrow bandwidths or domain-specific data, highlighting the need for approaches that generalize across diverse, high-resolution audio.

\section{Method}
\label{sec:pagestyle}

\subsection{Overview}
An overview of our approach is given in \autoref{fig:overview}. As in \cite{lahrichi2025qincodecneuralaudiocompression}, we design a three-step training process comprising (i) diffusion autoencoder training, (ii) offline quantization, and (iii) diffusion decoder finetuning.  Our approach operates entirely in the latent space $x_0$ of a pretrained Audio Autoencoder (AudioAE). First, we train a continuous diffusion autoencoder that relies on a diffusion decoder $D_{\theta}$ and a latent encoder $g_{\psi}$. The latent encoder maps latent vectors $x_0$ into compressed representations $z$ which are re-projected by a linear layer $f_{\phi}$ and used to condition $D_{\theta}$.
The decoder is a pretrained \acf{DiT}, finetuned using LoRA \cite{hu2021loralowrankadaptationlarge} adapters and trained together with $g_{\psi}$ and $f_{\phi}$, preserving generative priors while enforcing strong audio conditioning. Next, a neural quantizer is trained on the frozen compressed representation $z$ to obtain $z_q$, minimizing the distortion error. Finally, the diffusion decoder is finetuned by using the audio conditionings $f_{\phi}(z_q)$ instead of $f_{\phi}(z)$,  compensating for the information loss induced by quantization. 

\subsection{Pretraining}

\textit{AudioAE:} The latent vectors $x_0$ are derived from a pretrained low-compression audio autoencoder, yielding high-fidelity reconstructions in a
more compact and informative subspace.
Following prior work \cite{lahrichi2025qincodecneuralaudiocompression}, the AudioAE decoder is built upon the design of \cite{siuzdak2024vocosclosinggaptimedomain}, a GAN-based vocoder trained to predict STFT complex coefficients, and the encoder mirrors this structure.
This architecture preserves temporal resolution, reducing upsampling artifacts~\cite{DBLP:journals/corr/abs-2010-14356} and enabling efficient training with convolutions over uniformly sized sequences.

 \textit{\ac{DiT}:} $D_{\theta}$ is a pretrained text-to-audio \ac{DiT} 
 trained to denoise noisy latent vectors $x_0 + \sigma \epsilon$,  $\epsilon \sim \mathcal{N}(0, 1)$,  conditioned with a text encoder. It consists of sequential transformer blocks:  the first six are multi-modal blocks from \cite{esser2024scaling} operating on the \emph{audio} and \emph{text} modalities followed by six standard transformer blocks operating only on the \emph{audio} modality.
 
\begin{figure}[b!] 
    \centering
    \includegraphics[width=0.49\textwidth]{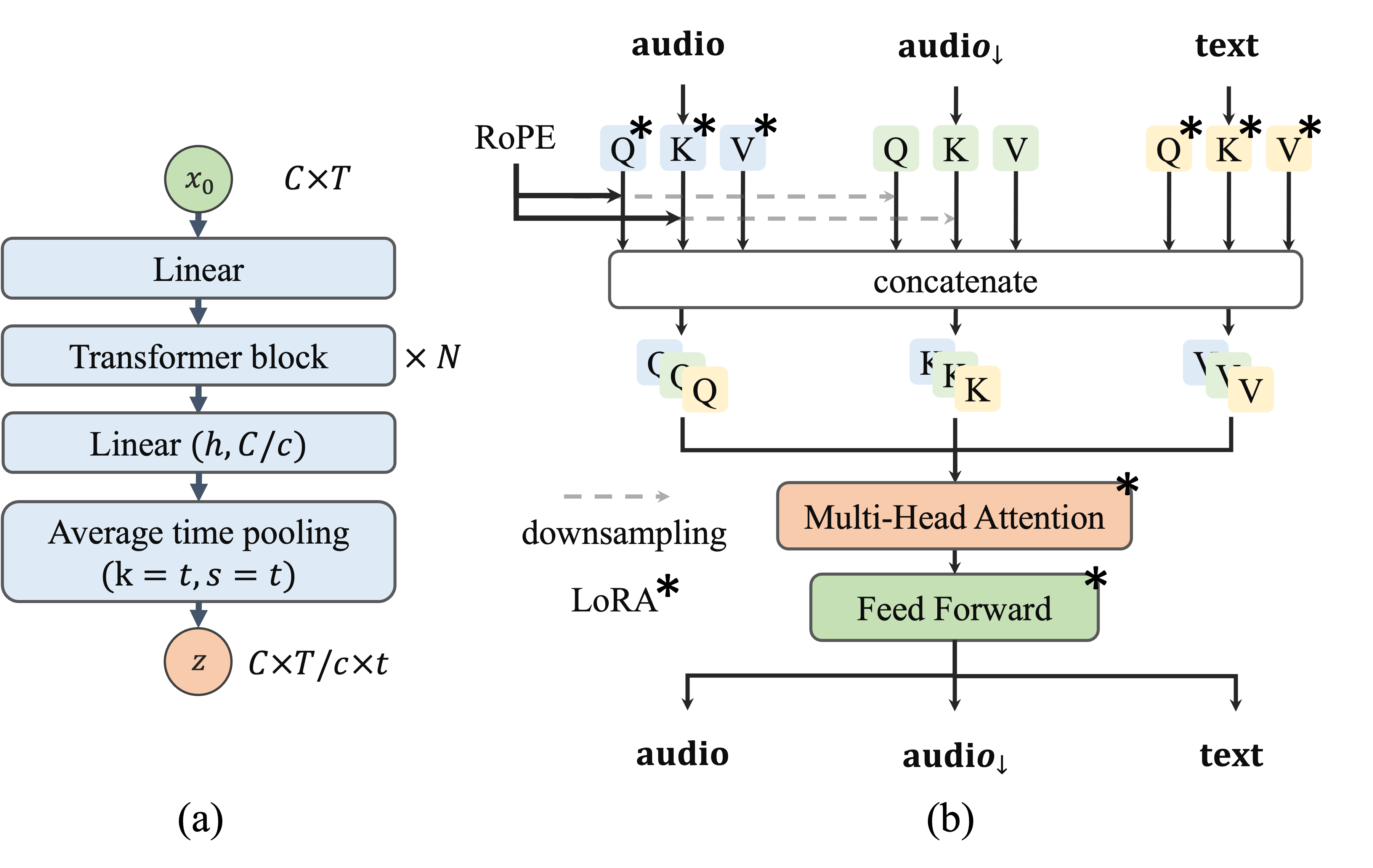}
    \caption{(a) Overview of the latent encoder architecture (b) Conditioning mechanism within the diffusion decoder.
    }
    \label{fig:architectures}
\end{figure}

\subsection{Diffusion Autoencoder training} 

\textit{Latent encoder:} The latent encoder further compresses the latent audio representations $x_0$ into $z$. The architecture of the latent encoder $g_{\psi}$ is provided in \autoref{fig:architectures}(a). The encoder downsamples the latent variables $x_0 \in \mathbb{R}^{C \times T}$ along frequency and time by factors $c$ and $t$, respectively. This is achieved through sequential transformer blocks, followed by a linear layer to reduce the dimensionality and an average pooling layer to downsample in time with kernel and stride $t$. The transformer blocks employ RoPE positional embeddings \cite{su2024roformer}, which will also be downsampled in the \ac{DiT} by selecting the central position of each temporal window.

\textit{Diffusion decoder:} Weights from $D_{\theta}$ are finetuned using LoRA adapters  \cite{hu2021loralowrankadaptationlarge} while audio conditioning is injected from the latent encoder via an additional linear layer $f_{\phi}$. 
The compressed audio conditioning is treated as a third modality termed \emph{audio$\downarrow$} similar to the image conditioning from \cite{batifol2025flux} and we initialize dedicated layers for its Q, K, V as depicted in \autoref{fig:architectures}(b). The RoPE for the audio conditioning \emph{audio$\downarrow$} are a decimated version of the RoPE used for $\emph{audio}$, so that we maintain the temporal alignment between the two modalities.

\subsection{Offline quantization}

As in \cite{lahrichi2025qincodecneuralaudiocompression}, the audio embeddings derived from our latent encoder $z = g_{\psi}(x_0)$ are quantized offline using the Qinco2 a neural quantizer \cite{ vallaeys2025qinco2vectorcompressionsearch}.
Qinco2 extends \ac{RVQ}, where continuous vectors are quantized through hierarchical codebooks. Instead of fixed codebooks, it employs neural networks to generate adaptive centroids conditioned on prior reconstructions, enabling finer modeling of residual distributions and capturing inter-codebook dependencies. The bitrate is determined by the number of codebooks $M$, codebook size $K$, and frame rate $f$: $M \times \log_2 K \times f$.

\subsection{Finetuning}

After training the codebooks, the latent embeddings $z$ are quantized into $z_q$, which are then re-injected in replacement of $z$ into the diffusion autoencoder pipeline. 
To mitigate the degradation induced by quantization, the LoRA weights of the diffusion decoder $D_{\theta}$ and the  projection layer $f_{\phi}$ of the transformer are finetuned. 

\section{EXPERIMENTS}

\subsection{Datasets}

All models are trained on a combination of four internal sound effect datasets, reaching $\sim$ 5000 hours, sampled at 48kHz and clipped to 5 seconds. 
Sound effects cover a wide range of audio types, including Foley sounds, environmental sounds, individual events, as well as music samples and background speech. 
Evaluation uses three datasets: Freesound Effects, BBC Sound Effects, and an internal studio-quality dataset, in order to assess performance on both public benchmarks and professional audio. We randomly sample 500 clips per dataset, each clipped or zero-padded to 5 seconds.

\begin{table*}
\centering
\small
\begin{tabular}{l|c|c|c|c|c|c|c|c|c}
\toprule[1.2pt]
Method & Variant & $D$ & Framerate & $R$ & FAD $\downarrow$ & FAD$_{\text{CLAP}}$ $\downarrow$ & KAD$_{\text{CLAP}}$ $\downarrow$ & CLAP$_{\text{audio}}$ $\uparrow$ & Si-SDR$\uparrow$ \\
\midrule
AudioAE          &    --    & 128 & 100 Hz  & 4   & 0.008         & 0.008          & 0.15          & 0.90          & 22.3           \\
\midrule
StableAudio Open &    --    & 64  & 21.5 Hz & 32  & 0.78          & 0.066          & 1.25          & \textbf{0.78} & 0.48           \\
\mymodel         &   $t=4$  & 64  & 25 Hz   & 30  & \textbf{0.48} & \textbf{0.038} & \textbf{0.57} & 0.76          & \textbf{3.21}  \\ 
\midrule
Music2Latent     &    --    & 64  & 11 Hz   & 64  & 1.28          & 0.168          & 3.29          & 0.69          & -10.5          \\
\mymodel         &   $t=9$  & 64  & 11 Hz   & 68  & \textbf{0.59} & \textbf{0.050} & \textbf{0.77} & \textbf{0.76} & \textbf{-2.40} \\
\midrule
\multirow{2}{*}{\mymodel} 
                 & $t=20$   & 64  & 5 Hz    & 150 & 0.76          & 0.059          & 0.92          & 0.71          & -8.80          \\
                 & $t=100$  & 64  & 1 Hz    & 750 & 0.64          & 0.059          & 0.89          & 0.73          & -27.7          \\
\bottomrule[1.2pt]
\end{tabular}
\caption{Performance in the continuous case of \mymodel{} vs. continuous audio compression baselines at equivalent compression rates $R$.}
\label{tab:continuous_baselines}
\end{table*}

\begin{table*}
\centering
\begin{tabular}{c|l|c|c|c|ccccccc}
\toprule[1.2pt] 
& Method       & kbps & Framerate & $M$ & FAD $\downarrow$ 
& FAD$_{\text{CLAP}}$ $\downarrow$ 
& KAD$_{\text{CLAP}}$ $\downarrow$
& CLAP$_{\text{audio}}$ $\uparrow$
& Si-SDR $\uparrow$ \\
\midrule
\multirow{3}{*}{\textbf{Low bitrates}} 
& DAC                & 1.7   & 86 Hz  & 2  & 3.24           & 0.108           & 1.71            & 0.63          & \textbf{-4.11} \\ 
& SemantiCodec       & 1.4   & 100 Hz & 1  & 1.79           & 0.136           & 4.93            & 0.60          & -31.8          \\ 
& \mymodel           & 1.32  & 11 Hz  & 12 & \textbf{0.55}  & \textbf{0.048}  & \textbf{0.728}  & \textbf{0.73} & -4.48          \\ 
\midrule
\multirow{3}{*}{\textbf{Ultra low bitrates}} 
& SemantiCodec       & 0.3125 & 25 Hz & 1  & 1.23           & 0.271           & 2.70            & 0.48          & -34.5          \\ 
& \mymodel           & 0.3    & 1 Hz  & 25 & \textbf{0.64}  & \textbf{0.052}  & \textbf{0.78}   & \textbf{0.71} & \textbf{-27.8} \\ 
& \mymodel           & 0.096  & 1 Hz  & 8  & 0.68           & 0.060           & 0.89            & 0.67          & -30.4          \\
\bottomrule[1.2pt] 
\end{tabular}
\caption{Performance in the discrete case of \mymodel{} vs. baseline audio codecs at equivalent bitrates.}
\label{tab:codec_baselines}
\end{table*}

\subsection{Training details}

\subsubsection{Pretraining}

\textit{AudioAE:} Input STFTs are computed using $n_\text{fft}=960$ and a hop size of $480$, yielding roughly 100 frames per second. The latent dimension is $C=128$. Training follows~\cite{kumar2023highfidelityaudiocompressionimproved}, reusing its discriminators, loss functions and optimization parameters. 

\textit{\ac{DiT}:} We adopt the EDM2~\cite{karras2022elucidatingdesignspacediffusionbased} parametrization and training strategy with text conditioning, using AdamW with a $1\mathrm{e}{-4}$ learning rate. Text conditioning is provided by a CLAP encoder~\cite{wu2024largescalecontrastivelanguageaudiopretraining} trained on our datasets.

\subsubsection{Three-step training}

\textit{Latent encoder:} 
We set the encoder depth by framerate, with lower rates requiring more complexity: 6 blocks at 25Hz, 10 at 11Hz, and 12 at 5Hz and 1Hz.
We set the frequency compression factor to $c=2$, yielding 64-dimensional representations, and choose $t$ according to the target frame rate.

\textit{Diffusion Decoder:} The DiT is finetuned with LoRA weights under the same EDM2 strategy, sampling $\sigma$ with a log normal distribution. To prioritize audio over text conditioning, we apply strong dropout (0.8) on the text embeddings, encouraging the model to rely mainly on audio conditioning. Our models are trained using the AdamW optimizer with a learning rate of $\mathrm{1e^{-4}}$. Each of them is trained across four A100 GPUs, with a batch size of $\mathrm{32}$. 

\textit{Offline quantization:} We follow the default parameters from Qinco2 \cite{vallaeys2025qinco2vectorcompressionsearch}, adjusting only the codebook size $K$ to match the task complexity. For 

high temporal compression rates ($t=100,20$), larger codebooks (12 bits vs.~10 bits) are used to reduce MSE in early quantization layers, as fewer frames encode more information and a larger vocabulary improves reconstruction. Each quantizer is trained with $M = 20$ codebooks, and a batch size of $8000$ vector frames.

\textit{Finetuning:} In the final finetuning stage, we replace $z$ with $z_q$ using $M$ codebooks, and continue training $f_\phi$ and the LoRA weights for 40 additional epochs. The choice of $M$ is determined by evaluating reconstructions with $z_q$ substituted for $z$ without decoder finetuning. As expected, fewer codebooks increase distortion and induce distributional drift. Empirically, we found that a total vocabulary size of about 100 bits is sufficient to retain most salient elements to stay close from the original sound. Accordingly, we finetune our models with $M=10$ for $K=10$ and $M=8$ for $K=12$. To stabilize training when replacing $z$ with $z_q$, we retain the original $z$ 10\% of the time, reducing the abrupt distributional shift.

\begin{figure}[b!]
    \centering
    \includegraphics[width=0.30\textwidth]{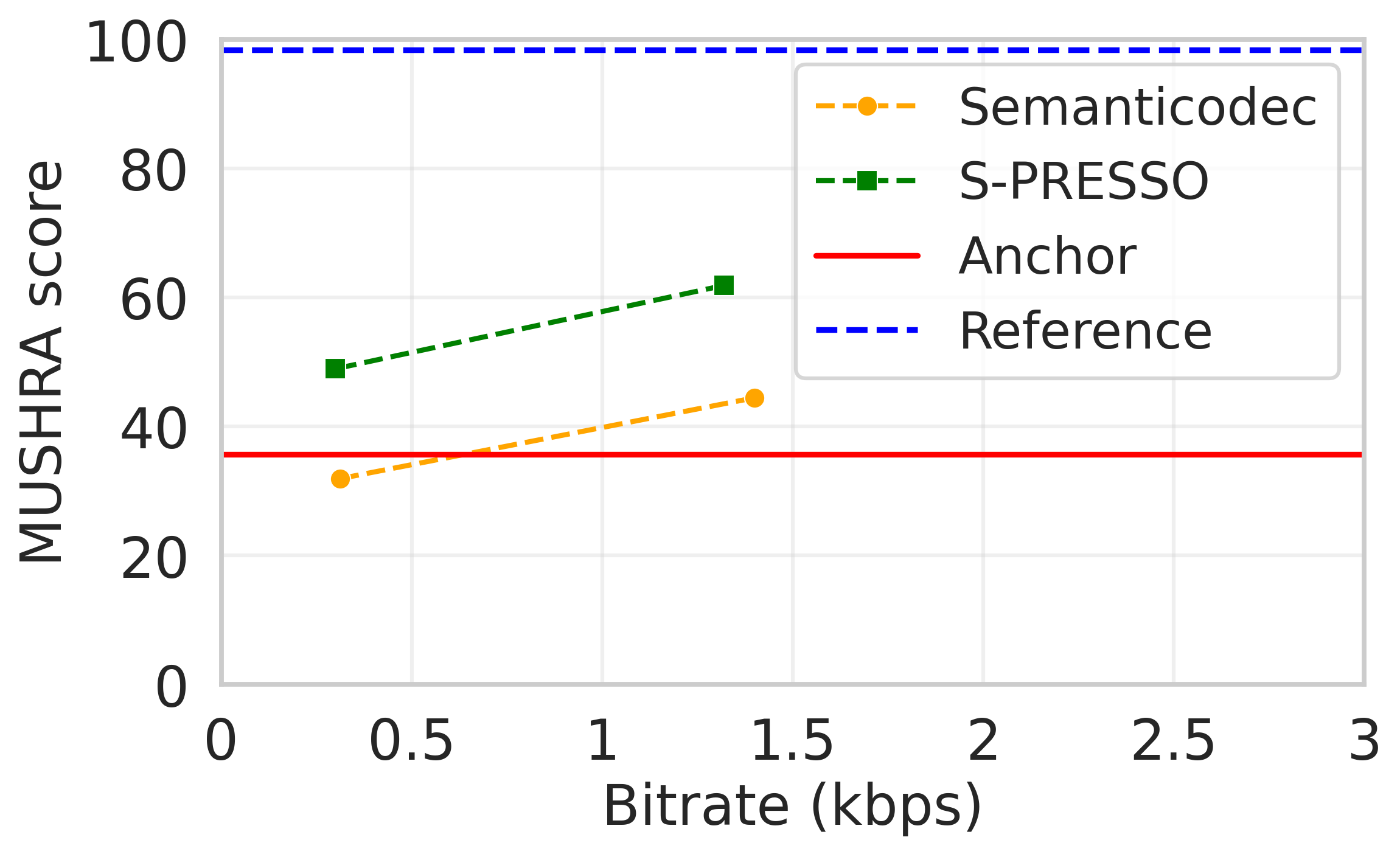}
    \caption{MUSHRA scores for \mymodel, SemantiCodec and a 3.5kHz low-pass anchor, evaluated at $\sim$ 1.35 kbps and $\sim$ 0.3 kbps.}
    \label{fig:mushra}
\end{figure}

\subsection{Baselines}

In the continuous case (step 1), we benchmark \mymodel{} against continuous baselines trained on high-quality audio: Stable Audio Open~\cite{evans2025stable} (44.1~kHz) and Music2Latent~\cite{pasini2024music2latent} (48~kHz). 
Music2Latent is trained exclusively on music, and is therefore not directly comparable but remains relevant since our evaluation sets spans background voices and music samples. For fairness, we train \mymodel{} with different downsampling factors $t$ to match the compression rates of the baselines. 
\autoref{tab:continuous_baselines} summarizes the latent dimensionality $D$, frame rate, and overall compression factor $R$ of each method. 
Additionally, we include the performance of our AudioAE, which serves as the upper bound for our models.

In the discrete case (step 3), the main baseline is SemantiCodec~\cite{Liu_2024}, a 16~kHz diffusion-based codec trained on general sounds at comparable bitrates. We evaluate both low bitrate (1–2~kbps) and ultra-low bitrate (100–300~bps) configurations. 
\autoref{tab:codec_baselines} summarizes the bitrate and frame rate of the compared methods. Finally, for consistency across models, audio clips are first resampled to each model’s native rate and then converted back to 48~kHz.

\subsection{Evaluation metrics}

Since our method relies on a generative decoder, we assess audio quality with the Fréchet Audio Distance (FAD) using VGGish and LAION-CLAP embeddings. We additionally report the Kernel Audio Distance (KAD)~\cite{chung2025kadfadeffectiveefficient} with LAION-CLAP embeddings, a distribution-free alternative to FAD which shows stronger correlation with human perception. 
We measure the reconstruction using the Si-SDR and the global acoustic similarity using the cosine distance between the CLAP audio embeddings.
For subjective evaluation, we conducted a MUSHRA test~\cite{Schoeffler2018webMUSHRAA} against SemantiCodec at low and ultra–low bitrates, using three samples per test set (foleys, music samples, ambiances) and 20 listeners (experts and non-experts) on headphones, who rated quality and similarity.
 
\begin{figure}[h]
    \includegraphics[width=0.48\textwidth]{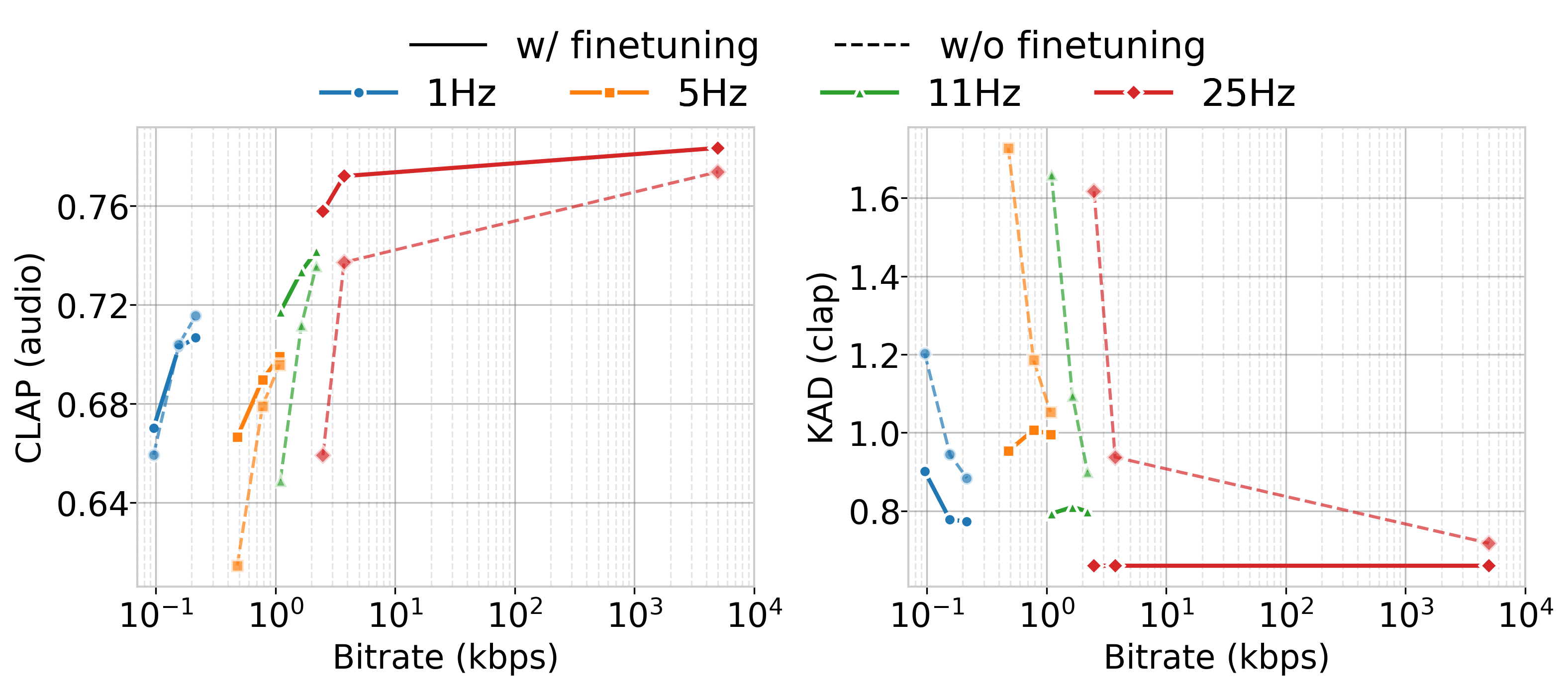}
    \caption{Evaluation of \mymodel{} at varying bitrates and framerates.}
    \label{fig:framerates}
\end{figure}

\section{RESULTS}
\label{sec:results}

For all experiments, we sample \mymodel{} reconstructions using the Heun solver~\cite{karras2022elucidatingdesignspacediffusionbased} with 64 steps and the default parameters.

\textit{Overall Performance:} 
As shown in the continuous compression benchmark \autoref{tab:continuous_baselines}, 
the AudioAE provides the upper bound on achievable quality. Relative to this reference, \mymodel{} consistently outperforms continuous baselines at comparable compression rates, achieving superior audio quality (FAD, KAD, Si-SDR) and acoustic similarity (CLAP), even at extreme compression rates (1 Hz). 
As the framerate increases, i.e. more temporal information is preserved, SI-SDR improves accordingly. By contrast, CLAP similarity remains remarkably stable across framerates.
This stability shows that reconstructions globally preserve acoustic category and source identity regardless of temporal downsampling.

\autoref{tab:codec_baselines} reports results in the discrete case. \mymodel{} outperforms Semanticodec across all metrics, with the latter’s 16~kHz bandwidth likely explaining its weaker FAD and KAD. These results confirm that bitrate does not constrain either perceptual quality or global similarity. Finally, as shown in \autoref{fig:mushra}, the MUSHRA test corroborates these findings, showing superior ratings for \mymodel{} at low and ultra–low bitrates. However, both models remain below the reference, partly because the variability of diffusion sampling complicates subjective judgments, forcing listeners to balance quality against similarity.

\textit{Impact of the bitrate:} In \autoref{fig:framerates}, we compare the performance of \mymodel{} in the discrete case for varying bitrates and framerates. 
The results further show that finetuning (step 3) consistently improves performance across bitrates, even though the model was not explicitly trained for variable bitrate settings. At a fixed framerate, more codebooks yield higher scores, reflecting finer residual modeling. Conversely, at a fixed bitrate, higher framerates perform better, highlighting the cost of framerate reduction.

\textit{Reconstruction diversity:} Empirically, at fixed compression rate, we observe more diversity with low frame rates.
Specifically, higher frame rates capture fine local structure, while lower frame rates (e.g., 1Hz, encoding a single vector) emphasize global information, leading to coarser reconstructions and increased variability across samples. This usually implies subtle differences in audio textures, high frequency details and background noise. Interestingly, we also observe that the model tends to replace background noise with other noise patterns, suggesting that it prioritizes preserving salient audio elements while freely re-synthesizing less critical components.

\vspace{-3pt}
\section{Conclusion}

We introduced \textbf{\mymodel}, a diffusion autoencoder for ultra–low bitrate compression of 48~kHz sound effects. Leveraging diffusion priors and strong multimodal conditioning, S-PRESSO achieves up to 750$\times$ compression while preserving perceptual quality. Our results suggest to shift the focus from strict fidelity to acoustic similarity enabling realistic and semantically consistent reconstructions even at 1Hz frame rates. Beyond surpassing continuous and discrete baselines, this work highlights the potential of generative models to redefine the limits of neural audio compression. While this work focused on audio quality, future efforts will make \mymodel{} more practical by speeding up inference time and extending it to general audio.




\newpage
\bibliographystyle{IEEEbib}
\bibliography{refs}

@inproceedings{siahkoohi2022ultralowbitratespeechcodingpretrained,
author = {Siahkoohi, Ali and Chinen, Michael and Denton, Tom and Kleijn, W. and Skoglund, Jan},
year = {2022},
month = {09},
pages = {4421-4425},
title = {Ultra-Low-Bitrate Speech Coding with Pretrained Transformers},
doi = {10.21437/Interspeech.2022-10988}
}

@article{xu2025mucodecultralowbitratemusic,
  title={Mucodec: Ultra low-bitrate music codec},
  author={Xu, Yaoxun and Chen, Hangting and Yu, Jianwei and Tan, Wei and Gu, Rongzhi and Lei, Shun and Lin, Zhiwei and Wu, Zhiyong},
  journal={arXiv preprint arXiv:2409.13216},
  year={2024}
}

@article{Liu_2024,
  title={Semanticodec: An ultra low bitrate semantic audio codec for general sound},
  author={Liu, Haohe and Xu, Xuenan and Yuan, Yi and Wu, Mengyue and Wang, Wenwu and Plumbley, Mark D},
  journal={IEEE Journal of Selected Topics in Signal Processing},
  year={2024},
  publisher={IEEE}
}

@article{karras2022elucidatingdesignspacediffusionbased,
  title={Elucidating the design space of diffusion-based generative models},
  author={Karras, Tero and Aittala, Miika and Aila, Timo and Laine, Samuli},
  journal={Advances in neural information processing systems},
  volume={35},
  pages={26565--26577},
  year={2022}
}

@article{wu2024largescalecontrastivelanguageaudiopretraining,
  title={Large-Scale Contrastive Language-Audio Pretraining with Feature Fusion and Keyword-to-Caption Augmentation},
  author={Yusong Wu and K. Chen and Tianyu Zhang and Yuchen Hui and Taylor Berg-Kirkpatrick and Shlomo Dubnov},
  journal={IEEE International Conference on Acoustics, Speech and Signal Processing (ICASSP)},
  year={2022},
  pages={1-5},
  url={https://api.semanticscholar.org/CorpusID:253510826}
}

@article{siuzdak2024vocosclosinggaptimedomain,
  title={Vocos: Closing the gap between time-domain and fourier-based neural vocoders for high-quality audio synthesis},
  author={Siuzdak, Hubert},
  journal={arXiv preprint arXiv:2306.00814},
  year={2023}
}

@article{DBLP:journals/corr/abs-2010-14356,
  author       = {Jordi Pons and
                  Santiago Pascual and
                  Giulio Cengarle and
                  Joan Serr{\`{a}}},
  title        = {Upsampling artifacts in neural audio synthesis},
  journal      = {CoRR},
  volume       = {},
  year         = {2020},
  url          = {https://arxiv.org/abs/2010.14356},
  eprinttype    = {arXiv},
  eprint       = {2010.14356},
  bibsource    = {dblp computer science bibliography, https://dblp.org}
}

@article{kumar2023highfidelityaudiocompressionimproved,
  title={High-fidelity audio compression with improved rvqgan},
  author={Kumar, Rithesh and Seetharaman, Prem and Luebs, Alejandro and Kumar, Ishaan and Kumar, Kundan},
  journal={Advances in Neural Information Processing Systems},
  volume={36},
  pages={27980--27993},
  year={2023}
}

@article{chung2025kadfadeffectiveefficient,
  title={KAD: No More FAD! An Effective and Efficient Evaluation Metric for Audio Generation},
  author={Chung, Yoonjin and Eu, Pilsun and Lee, Junwon and Choi, Keunwoo and Nam, Juhan and Chon, Ben Sangbae},
  journal={arXiv preprint arXiv:2502.15602},
  year={2025}
}

@inproceedings{careil2024imagecompressionperfectrealism,
  title={Towards image compression with perfect realism at ultra-low bitrates},
  author={Careil, Marlene and Muckley, Matthew J and Verbeek, Jakob and Lathuili{\`e}re, St{\'e}phane},
  booktitle={The Twelfth International Conference on Learning Representations},
  year={2023}
}

@article{vallaeys2025qinco2vectorcompressionsearch,
  title={Qinco2: Vector compression and search with improved implicit neural codebooks},
  author={Vallaeys, Th{\'e}ophane and Muckley, Matthew and Verbeek, Jakob and Douze, Matthijs},
  journal={arXiv preprint arXiv:2501.03078},
  year={2025}
}

@article{hu2021loralowrankadaptationlarge,
  title={Lora: Low-rank adaptation of large language models.},
  author={Hu, Edward J and Shen, Yelong and Wallis, Phillip and Allen-Zhu, Zeyuan and Li, Yuanzhi and Wang, Shean and Wang, Lu and Chen, Weizhu and others},
  journal={ICLR},
  volume={1},
  number={2},
  pages={3},
  year={2022}
}

@inproceedings{lahrichi2025qincodecneuralaudiocompression,
  title={{QINCODEC:} Neural Audio Compression with Implicit Neural Codebooks},
  author       = {Zineb Lahrichi and
                  Ga{\"{e}}tan Hadjeres and
                  Ga{\"{e}}l Richard and
                  Geoffroy Peeters},
  booktitle={European Signal Processing Conference (EUSIPCO)},
  year={2025},

}

@article{su2024roformer,
  title={Roformer: Enhanced transformer with rotary position embedding},
  author={Su, Jianlin and Ahmed, Murtadha and Lu, Yu and Pan, Shengfeng and Bo, Wen and Liu, Yunfeng},
  journal={Neurocomputing},
  volume={568},
  pages={127063},
  year={2024},
  publisher={Elsevier}
}

@article{défossez2022highfidelityneuralaudio,
  title={High fidelity neural audio compression},
  author={D{\'e}fossez, Alexandre and Copet, Jade and Synnaeve, Gabriel and Adi, Yossi},
  journal={arXiv preprint arXiv:2210.13438},
  year={2022}
}

@inproceedings{kleijn2018wavenet,
  title={Wavenet based low rate speech coding},
  author={Kleijn, W Bastiaan and Lim, Felicia SC and Luebs, Alejandro and Skoglund, Jan and Stimberg, Florian and Wang, Quan and Walters, Thomas C},
  booktitle={IEEE international conference on acoustics, speech and signal processing (ICASSP)},
  pages={676--680},
  year={2018},

}

@article{batifol2025flux,
  title={FLUX. 1 Kontext: Flow Matching for In-Context Image Generation and Editing in Latent Space},
  author={Batifol, Stephen and Blattmann, Andreas and Boesel, Frederic and Consul, Saksham and Diagne, Cyril and Dockhorn, Tim and English, Jack and English, Zion and Esser, Patrick and Kulal, Sumith and others},
  journal={arXiv e-prints},
  pages={arXiv--2506},
  year={2025}
}

@article{zeghidour2021soundstream,
  title={Soundstream: An end-to-end neural audio codec},
  author={Zeghidour, Neil and Luebs, Alejandro and Omran, Ahmed and Skoglund, Jan and Tagliasacchi, Marco},
  journal={IEEE/ACM Transactions on Audio, Speech, and Language Processing},
  volume={30},
  pages={495--507},
  year={2021},
  publisher={IEEE}
}

@inproceedings{evans2025stable,
  title={Stable audio open},
  author={Evans, Zach and Parker, Julian D and Carr, CJ and Zukowski, Zack and Taylor, Josiah and Pons, Jordi},
  booktitle={IEEE International Conference on Acoustics, Speech and Signal Processing (ICASSP)},
  pages={1--5},
  year={2025},

}

@article{pasini2024music2latent,
  title={Music2latent: Consistency autoencoders for latent audio compression},
  author={Pasini, Marco and Lattner, Stefan and Fazekas, George},
  journal={arXiv preprint arXiv:2408.06500},
  year={2024}
}

@article{Schoeffler2018webMUSHRAA,
  title={webMUSHRA — A Comprehensive Framework for Web-based Listening Tests},
  author={Michael Schoeffler and Sarah Bartoschek and Fabian-Robert St{\"o}ter and Marlene Roess and Susanne Westphal and Bernd Edler and J{\"u}rgen Herre},
  journal={Journal of open research software},
  year={2018},
  volume={6},
}

@inproceedings{casanova2025low,
  title={Low frame-rate speech codec: a codec designed for fast high-quality speech LLM training and inference},
  author={Casanova, Edresson and Langman, Ryan and Neekhara, Paarth and Hussain, Shehzeen and Li, Jason and Ghosh, Subhankar and Juki{\'c}, Ante and Lee, Sang-gil},
  booktitle={IEEE International Conference on Acoustics, Speech and Signal Processing (ICASSP)},
  pages={1--5},
  year={2025},

}

@article{li2025dualcodec,
  title={Dualcodec: A low-frame-rate, semantically-enhanced neural audio codec for speech generation},
  author={Li, Jiaqi and Lin, Xiaolong and Li, Zhekai and Huang, Shixi and Wang, Yuancheng and Wang, Chaoren and Zhan, Zhenpeng and Wu, Zhizheng},
  journal={arXiv preprint arXiv:2505.13000},
  year={2025}
}

@inproceedings{esser2024scaling,
  title={Scaling rectified flow transformers for high-resolution image synthesis},
  author={Esser, Patrick and Kulal, Sumith and Blattmann, Andreas and Entezari, Rahim and M{\"u}ller, Jonas and Saini, Harry and Levi, Yam and Lorenz, Dominik and Sauer, Axel and Boesel, Frederic and others},
  booktitle={Forty-first international conference on machine learning},
  year={2024}
}

\end{document}